\documentclass[10pt,twocolumn,letterpaper]{article}

\usepackage[pagenumbers]{cvpr} 

\usepackage{multirow}
\usepackage{tabularx}
\usepackage{booktabs}
\usepackage{adjustbox}
\usepackage{booktabs}
\usepackage{multirow}
\usepackage{graphicx}   
\definecolor{cvprblue}{rgb}{0.21,0.49,0.74}

\definecolor{red}{rgb}{1.0,0.0,0.0}

\usepackage[pagebackref,breaklinks,colorlinks,allcolors=cvprblue]{hyperref}

\def\paperID{10567} 
\def\confName{CVPR}
\def\confYear{2026}

\def\ttt{MatMart}

\title{\ttt: Material Reconstruction of 3D Objects via Diffusion}

\author{
  Xiuchao Wu$^{1,3*}$ \quad Pengfei Zhu$^{2,3*}$ \\ \quad Jiangjing Lyu$^{3\dagger}$ \quad Xinguo Liu$^{1}$ \quad Jie Guo$^{2}$ \quad Yanwen Guo$^{2}$ \quad Weiwei Xu$^{1}$ \quad Chengfei Lyu$^{3\ddagger}$\\
  $^{1}$Zhejiang University \quad $^{2}$Nanjing University \quad $^{3}$Alibaba Group
}

\begin{document}


\twocolumn[{%
\renewcommand\twocolumn[1][]{#1}%
\maketitle
\begin{center}
    \centering
    \includegraphics[width=\textwidth]{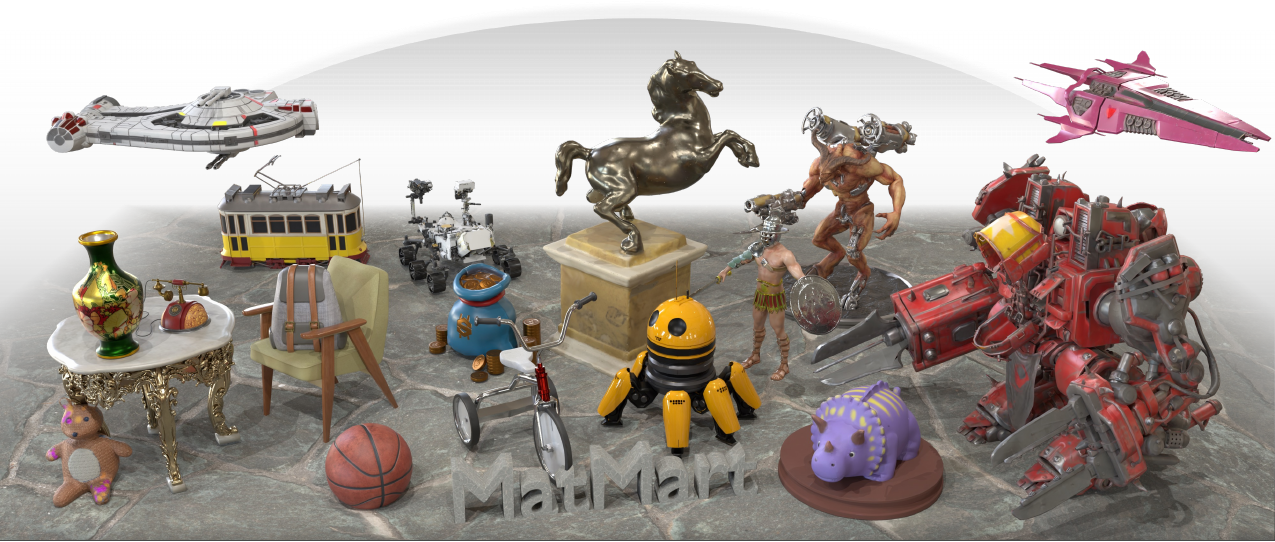}
    \captionof{figure}{\textbf{Reconstructed results in single-view input:} Our method reconstructs high-quality materials across various types of objects.}
\end{center}%
\vspace{1em}
}]

\renewcommand{\thefootnote}{} 
\footnotetext{$^*$Equal contribution \quad $\dagger$Project leader \quad $\ddagger$Corresponding author.}

\begin{abstract}

Applying diffusion models to physically-based material estimation and generation has recently gained prominence. In this paper, we propose \ttt, a novel material reconstruction framework for 3D objects, offering the following advantages. First, \ttt\ adopts a two-stage reconstruction, starting with accurate material prediction from inputs and followed by prior-guided material generation for unobserved views, yielding high-fidelity results. Second, by utilizing progressive inference alongside the proposed view-material cross-attention (VMCA), \ttt\ enables reconstruction from an arbitrary number of input images, demonstrating strong scalability and flexibility. Finally, \ttt\ achieves both material prediction and generation capabilities through end-to-end optimization of a single diffusion model, without relying on additional pre-trained models, thereby exhibiting enhanced stability across various types of objects. Extensive experiments demonstrate that \ttt\ achieves superior performance in material reconstruction compared to existing methods.

\end{abstract}    
\section{Introduction}
\label{sec:intro}

Recovering materials of a 3D object from RGB images has long been a challenge in computer vision and graphics.
Previous methods reconstruct material properties through differentiable rendering~\cite{Munkberg_2022_CVPR, hasselgren2022shape, jin2023tensoir, liang2024gs}, which typically involves per-object optimization and requires a sufficient number of captured images, consequently leading to low efficiency and high instability.
Recently, approaches leveraging diffusion models for material prediction and generation have emerged~\cite{li2024idarb,chen2024intrinsicanything,yu2024texgen,luo2024intrinsicdiffusion, zeng2024rgb,he2025materialmvp,huang2025material,kocsis2024intrinsic}, providing new perspectives for inverse material decomposition.

However, there remain challenges in building a material reconstruction framework with existing methods. 
First, reconstructing materials with high-fidelity requires both accurate material estimation and thorough preservation of details from the input images. Unfortunately, rich textures, especially meaningful ones such as text and logos present on objects, remain difficult to faithfully reproduce with current generative models.
Second, to be practical, the framework should possess scalability and be capable of handling varying numbers of input images with high-resolution. However, this is often constrained by the framework design and GPU memory.
Finally, it is advisable for the framework to avoid dependencies on outputs from other pre-trained models or on the use of multiple models, as these can increase the complexity of training and deployment while potentially diminishing the stability of the method.

To address the above challenges, we propose \ttt, a diffusion-based framework for high-fidelity and scalable material reconstruction, which operates in two stages. In the first stage, \ttt\ accurately estimates material properties from RGB inputs and bakes the results into UV space. In the second stage, \ttt\ adaptively selects views based on UV texel coverage, then alternates between prior-guided generation and texture baking for occluded and unobserved regions. This two-stage reconstruction can preserve the details from inputs as much as possible and fully utilize the latest prior information for generation, ultimately achieving highly faithful material recovery. To handle the potentially diverse quantities of input images present in various cases, we adopt a progressive inference strategy for both estimation and generation. As such inference does not allow cross-view attention across all views, we propose view-material cross-attention (VMCA) to ensure multi-view consistency. 
With this inference strategy, our framework supports prediction and generation from arbitrary numbers of inputs and higher resolutions (Fig.~\ref{fig:resolution}), thereby demonstrating strong scalability. Finally, in contrast to existing approaches that involve multiple models or rely on pre-trained models, \ttt\ unifies material prediction and generation within a single, jointly trainable model. Such a unified architecture greatly reduces the complexity of training and deployment, while also improves the stability.

In summary, our main contributions are as follows:
\begin{enumerate}[label=\textbullet,leftmargin=14pt]

\item We propose a novel two-stage framework for material reconstruction. Combining progressive material prediction with prior-guided generation, our framework recovers accurate PBR materials, thereby achieving highly faithful renderings across various types of objects.

\item By integrating the proposed view-material cross-attention (VMCA) into progressive inference, we achieve consistent outputs with reduced memory consumption, allowing our framework to recover materials from a single to arbitrary numbers of high-resolution inputs, thereby exhibiting strong scalability and flexibility.

\item Material prediction and generation are achieved using a unified architecture that operates with a single diffusion model and allows for end-to-end optimization.

\end{enumerate}
\begin{figure*}
\centering
\includegraphics[width=\linewidth]{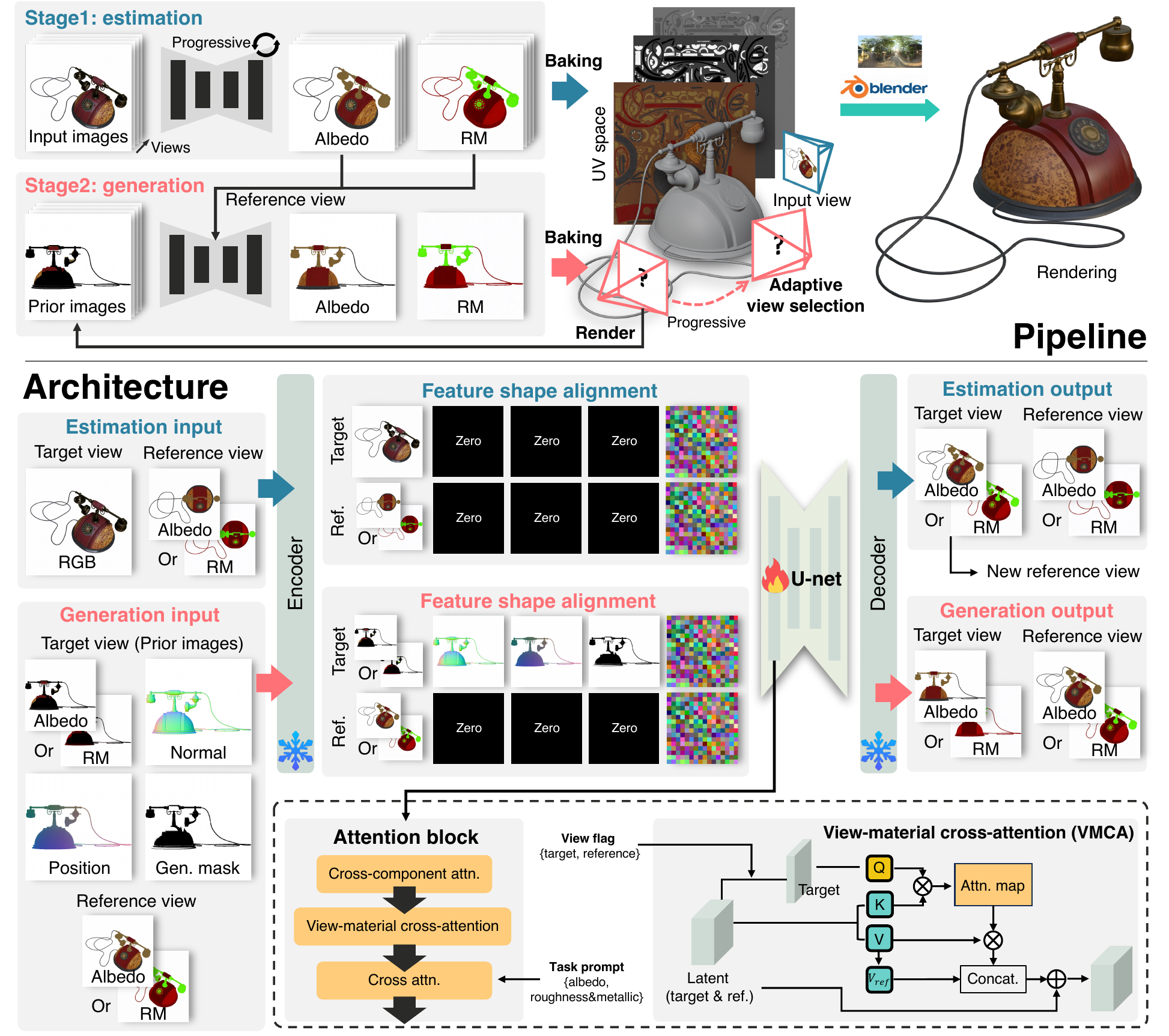}
\caption{\textbf{Method overview.} Our framework, \ttt, divides the material reconstruction task into two stages. In the first stage, progressive material prediction is performed on the input images, and the predicted results are baked into the UV space. In the second stage, prior-guided material generation and texture baking are alternately conducted for unobserved and occluded regions. Both prediction and generation tasks are unified within a single diffusion model and can be accomplished through end-to-end optimization.}
\label{fig:pipeline} 
\end{figure*}

\section{Related Work}
\label{sec:related_work}

\subsection{Optimization-based Material Reconstruction}
3D material reconstruction has been evolving for decades. 
The early approaches capture extensive reflectance samples using specialized devices such as gonioreflectometers~\cite{matusik2003data,dana1999reflectance,ward1992measuring} and fit BRDF parameters with high accuracy but significant acquisition cost. 
Recently, inverse rendering frameworks have provided end-to-end optimization of material, allowing high-quality reconstructions from image supervision alone~\cite{hasselgren2022shape,Munkberg_2022_CVPR,gao2019deep}.
Implicit representations have been introduced to jointly encode geometry and spatially continuous material based on neural fields and volumetric rendering, enabling high-fidelity rendering~\cite{liu2023nero,yao2022neilf,boss2021nerd,jin2023tensoir,li2024tensosdf}. 
Beyond implicit modeling, recent extensions of 3D Gaussian Splatting (3DGS) techniques attach material attributes to Gaussian primitives for fast material reconstruction~\cite{lai2025glossygs,Du_2025_ICCV,liang2024gs}. 
While most of these methods require large numbers of samples to optimize material parameters, diffusion-based models can generate new samples for optimization~\cite{jin2024neural,zhang2024dreammat} or directly predict high-quality, physically-consistent material maps~\cite{litman2025materialfusion,zeng2024rgb,chen2024intrinsicanything,li2024idarb,luo2024intrinsicdiffusion}, thereby enhancing reconstruction capabilities, marking a new stage in data-driven material reconstruction.

\subsection{Learning-based Material Diffusion}

With the advent of deep learning, convolutional and transformer-based networks began predicting material parameters directly from single or multi-view images~\cite{li2018learning,deschaintre2018single}. 
Recent advances in diffusion-based generation have produced a diverse set of methods for synthesizing textures and physically based rendering (PBR) materials, yet differences in how multi-view consistency is handled remain a key distinction. In texture generation, most approaches choose to either progressively predict views in sequence—explicitly enforcing cross-view consistency through conditioning~\cite{richardson2023texture,chen2023text2tex,huo2024texgen,poole2022dreamfusion}—or synthesize all views jointly using cross-attention to ensure coherent global structure~\cite{lu2025genesistex2,feng2025romantex,liang2025UnitTEX}. 
To achieve more realistic rendering effects, recent work leverages similar principles for full PBR material creation: some works adopt a two-stage paradigm, first producing photorealistic renderings through pretrained diffusion models and then recovering material parameters via inverse rendering pipelines~\cite{youwang2024paint,chen2023fantasia3d,zhang2024dreammat,qiu2024richdreamer,liu2024unidream} or learned prediction networks~\cite{zhang2024mapa,huang2025material}. While these techniques achieve impressive visual fidelity, most rely on complex multi-stage optimization, which can limit scalability and stability. Other approaches exploit geometric cues and conditions to directly generate multi-view material maps~\cite{he2025materialmvp,zhang2024clay}. However, these outputs, although semantically consistent, fail to maintain full fidelity to the conditions. These limitations motivate methods that unify material and texture generation within a high-faithfulness, end-to-end reconstruction framework.

\section{Method}
\label{sec:method}

\ttt\ is a diffusion-based material reconstruction framework, it aims to reconstruct high-quality PBR materials (\ie, albedo ${\bf A}$, roughness and metallic ${\bf RM}$) for 3D objects from input RGB images. As shown in Fig.~\ref{fig:pipeline}, this framework operates in two main stages. First, we perform progressive inference with the proposed view-material cross-attention to estimate materials for all input images, resulting in globally consistent predictions (Sec.\ref{sec:estimation}). Subsequently, we leverage the predicted materials from the first stage as priors to progressively guide the material generation process (Sec.~\ref{sec:generation}). Both prediction and generation tasks are unified within a single diffusion model, and the relevant training and inference details are presented in Sec.~\ref{sec:details}.

\subsection{Progressive Material Estimation}
\label{sec:estimation}

Given a set of RGB images, our goal is to estimate their material properties using image-based diffusion and to produce consistent results across the set. However, achieving this with existing techniques~\cite{li2024idarb, wu2025stableintrinsic} presents limitations. Specifically, to enforce multi-view consistency, these approaches employ cross-view attention, which requires concatenating all input images. This results in a space complexity of $O(N^2)$, which makes it challenging to handle dense or high-resolution inputs, ultimately limiting the scalability and reconstruction quality of the framework.
To address this problem, we adopt a progressive inference strategy, sequentially feeding input images into the network for estimation, and propose view-material cross-attention tailored to this process to enforce multi-view consistency.

\begin{figure}
\centering
\includegraphics[width=\linewidth]{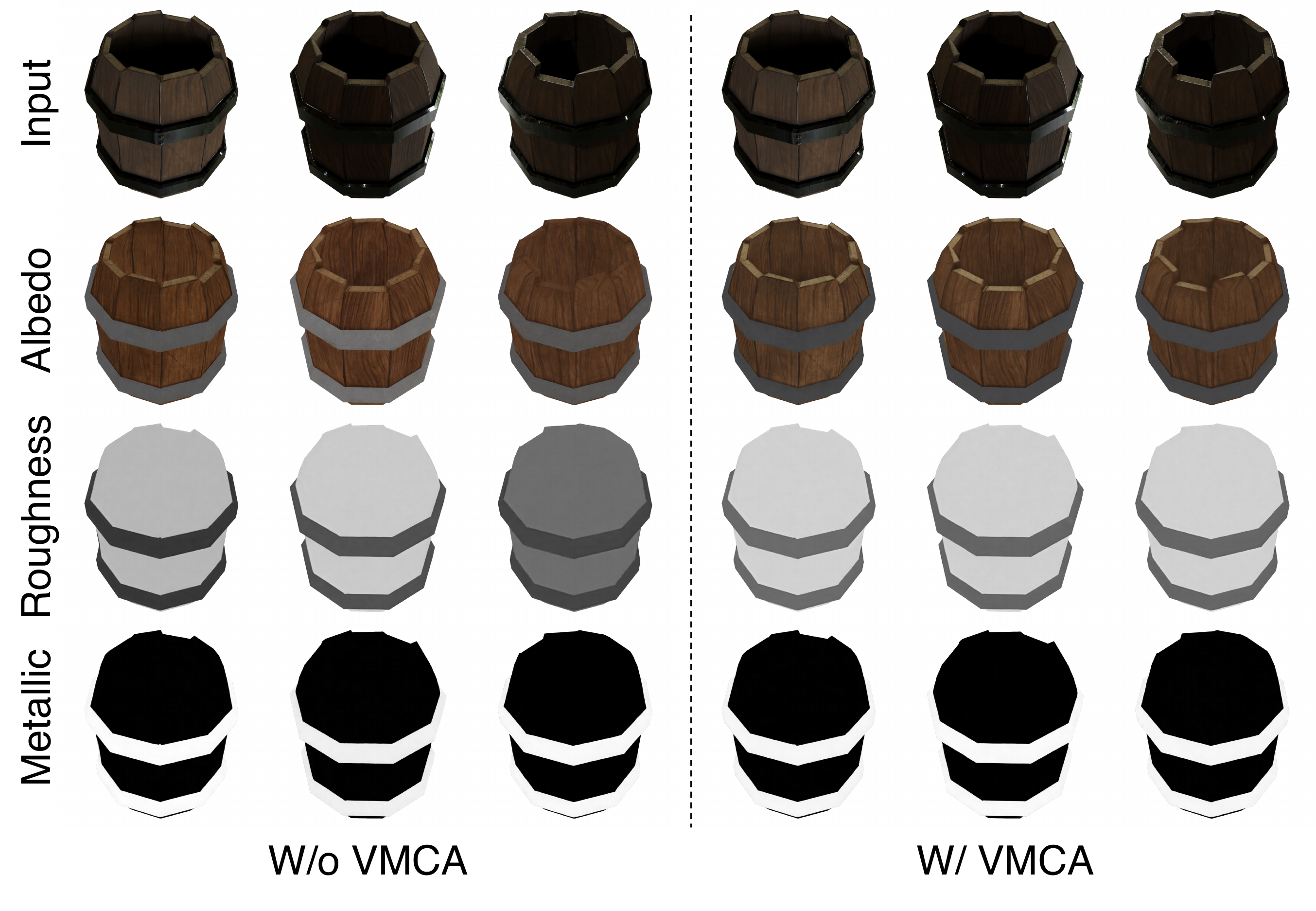}
\caption{\textbf{The effect of VMCA on predicted results.} VMCA improves the consistency for progressive material estimation.}.
\label{fig:vmca_pred} 
\end{figure}

\paragraph{View-material Cross-attention (VMCA).} 
As progressive inference breaks the information exchange among the inputs, we incorporate a reference view for each inference round. This view is the predicted result from the previous round and serves to establish connections among images inferred in different rounds. Therefore, there are two types of input for each inference: the target view to be estimated, and the reference view containing material information. To build the connection, we perform view-material cross-attention in the latent space between the target view and the reference view, thereby achieving consistent outputs across the views (Fig.~\ref{fig:vmca_pred}). This process can be formulated as:
\begin{align}
\mathbf{Z} = \left({\operatorname{Softmax}(\frac{\mathbf{Q}_\mathrm{Tgt}\cdot\mathbf{K}_\mathrm{Tgt+Ref}^T}{\sqrt{d}})}\cdot\mathbf{V}_\mathrm{Tgt+Ref}\right) \oplus \mathbf{V}_\mathrm{Ref},
\label{equ:vm}
\end{align}

where $\mathrm{Tgt}$ and $\mathrm{Ref}$ denote the indices for the target and reference view, respectively. $\mathbf{Q}$, $\mathbf{K}$ and $\mathbf{V}$ are the Query, Key and Value features. $\oplus$ denotes concatenation along the sequence dimension, and $\mathbf{Z}$ represents the latent features of target and reference views. As illustrated by Eq.~(\ref{equ:vm}), since the reference view is not expected to be influenced by the target view, it is not used as the Query but is instead directly output as $\mathbf{V}_\mathrm{Ref}$. As the numbers of target and reference views are fixed during inference and independent of the total number of input images, the overall space complexity of view-material cross-attention is $O(1)$.  This property allows the inference of higher resolutions and an unrestricted number of inputs under limited hardware resources.

\subsection{Prior-guided Material Generation}
\label{sec:generation}
\begin{figure*}
\centering
\includegraphics[width=\linewidth]{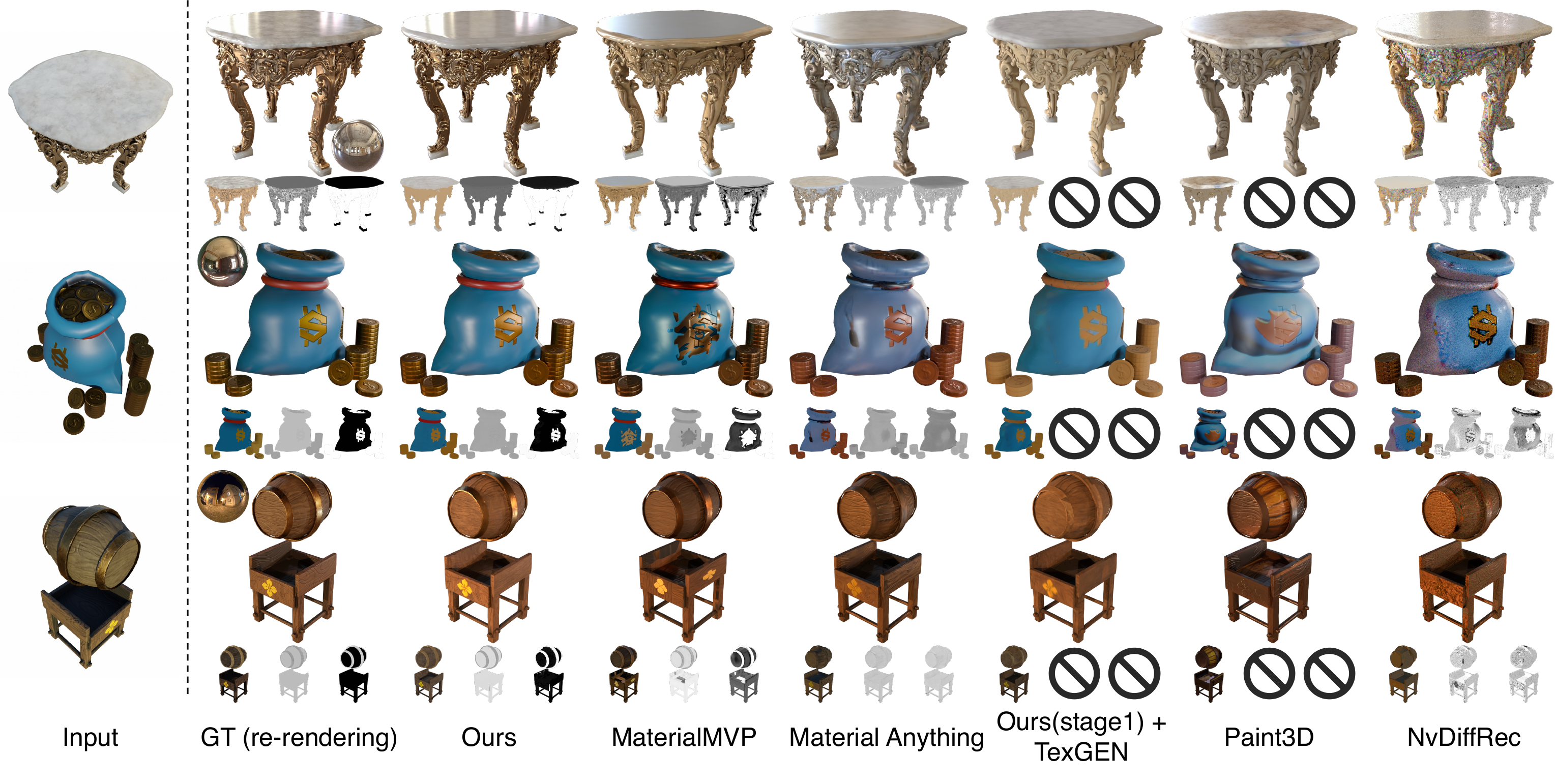}
\caption{\textbf{Qualitative comparison of single-view input.} Our method recovers more accurate materials across various types of objects, thereby achieving rendering results that are more consistent with the ground truth. From left to right are albedo, roughness, and metallic.}
\label{fig:com-single} 
\end{figure*}

After estimation, all predicted results are baked into UV space. However, in the case of single-view or sparse-view inputs, the baked UV space materials exhibit missing regions in occluded or unobserved parts, and therefore require further completion. Methods such as TEXGen~\cite{yu2024texgen} generate materials directly in UV space. We observe that such approaches currently struggle to produce high-quality results (Fig.~\ref{fig:com-single}), which may be attributed to the weak semantic information in UV space. Consequently, we opt to generate materials in view space, which has more semantic information. To obtain appropriate viewpoints for generation, we adaptively select views based on geometric information and the texel coverage in UV space. 
For each selected view, we project the UV space materials into the view space to obtain material priors (\ie, albedo, roughness and metallic). As illustrated in the bottom left of Fig.~\ref{fig:pipeline}, the material priors contain partial material information, which helps the network generate more consistent results (Fig.~\ref{fig:mat}). For regions requiring generation, we indicate them with a generation mask. In addition, we render geometric priors, including normal and point position, from the known geometry to assist the generation process. These prior images are used as inputs to the network to guide the generation process.

\paragraph{Adaptive View Selection.}

We select 6 views distributed along the coordinate axes as base views, since they are typically the most informative views. Then, we pick another $N$ views using an adaptive greedy strategy according to the texel coverage, which is obtained by projecting the selected views onto UV space.
Specifically, we uniformly sample 300 candidate viewpoints over the sphere and evaluate each of them. In each iteration, the one that yields the maximal improvement in UV space texel coverage is selected. This process is terminated when the target texel coverage $\rho$ is achieved or the number of selected views reaches $N$. We empirically set $\rho = 0.95$ and $N = 10$. 
Furthermore, to provide as much material priors as possible for each generation round, we sort the selected $N + 6$ views based on their generation masks. As a result, views containing fewer pixels to be generated are given higher priority for generation.

\paragraph{Alternating Baking and Group-wise Generation.}
Providing latest material priors for each generation step can make full use of existing information. Therefore, we perform material generation and baking alternately.
The baking process is similar to \cite{hunyuan3d2025hunyuan3d}, with modifications to work in a progressive manner. For each generated view space material, we project it onto UV space to obtain a texture map $\mathbf{T}'$, which is then used to update the existing texture map $\mathbf{T}$:
\begin{align}
\mathbf{T} &\leftarrow \frac{\mathbf{T}'\cdot \mathbf{W}' + \mathbf{T} \cdot \mathbf{W}}{\mathbf{W}' + \mathbf{W}}, \quad\mathbf{W'} = \mathbf{S'}^\lambda \\
\mathbf{W} &\leftarrow \mathbf{W}' + \mathbf{W},
\label{equ:update_texture}
\end{align}
where $\mathbf{W}$ and $\mathbf{W}'$ denote the blending weight for texture map $\mathbf{T}$ and $\mathbf {T}'$ respectively. $\mathbf{S}'$ represents the cosine similarity between the normal and the inverse forward axis of camera, which serves to reduce the material contribution at grazing angles. $\lambda$ is used to modulate the effect of $\mathbf{S}'$ (we follow \cite{hunyuan3d2025hunyuan3d} and set it to $6$). For generation, although progressive inference reduces memory consumption, we found that generating for only a single target view at a time is inefficient. Therefore, we choose to infer a group of target views at each step. The group size can be determined according to the memory capacity and GPU utilization (we set it to 3 in our experiments). In most of our experiments, we observe that the choice of group size does not affect the reconstruction quality. Additionally, we use the materials predicted in stage 1 as the reference view and employ VMCA to further enhance the consistency across groups.

\paragraph{Framework Discussion.}
\label{para:discuss}
Considering that Material Anything~\cite{huang2025material} similarly adopts a progressive generation strategy, we herein discuss the differences. Material Anything employs a pre-trained diffusion model to generate RGB images for all selected views, and then predicts their materials. In contrast, our method first predicts the materials of the input images and subsequently generates materials.  
Compared to generating RGB images, generating PBR materials is inherently simpler, as it avoids modeling intricate effects such as surface reflections and illumination variations. Our method does not rely on other pre-trained models, thereby avoids the instability introduced by such models. Furthermore, there may be domain differences between the generated RGB data and the training data of the material prediction model, which can reduce the model’s prediction accuracy and ultimately degrade rendering quality (Fig.~\ref{fig:com-single}).

\subsection{Implementation Details}
\label{sec:details}

\paragraph{Unified Architecture.}
Benefiting from our two-stage framework design that performs material prediction followed by material generation, the outputs of both tasks are unified as PBR materials. And both tasks leverage VMCA to enhance output consistency under progressive inference. These commonalities allow us to integrate both tasks within a unified architecture, thus simplifying the training and deployment process. Specifically, we employ a single diffusion model, enabling it to have capabilities for both material prediction and generation through end-to-end optimization. To align the shape of input tensors for both tasks, we pad the missing inputs with zeros in feature space. This also serves as an indicator for the model to distinguish between prediction and generation tasks. For the structural details, it is based on a pre-trained Stable Diffusion~\cite{rombach2022high} model. Each attention block consists of three attentions: a cross-component attention for information exchange between $\mathbf{A}$ and $\mathbf{RM}$, a view-material cross-attention to enhance consistency, and a cross-attention for controlling the type of output materials ($\mathbf{A}$ or $\mathbf{RM}$) by text prompt.

\paragraph{Training.} We alternately optimize our model between the prediction and generation tasks using 16 NVIDIA H20 GPU cards. In each iteration of training, we use two target views and one reference view, where the reference view here is set to the ground truth material. For efficiency, we use depth-based warping to obtain the material priors and generation masks during the optimization of generation task. As the initial view in progressive material estimation does not have a reference view, we additionally include the training for the prediction task without a reference view. The training dataset is the same as that used in IDArb~\cite{li2024idarb}, which consists of ABO~\cite{collins2022abo}, G-Objaverse~\cite{qiu2024richdreamer}, and Arb-Objaverse.
We start by training for 20K steps on $256 \times 256$ images with a batch size of 16, and then further train 50K steps on $512 \times 512$ images with a batch size of 4.
We use AdamW~\cite{loshchilov2017decoupled} for optimization, taking v-prediction~\cite{salimans2022progressive} as the target and assigning a learning rate of $1\times 10^{-4}$.

\begin{figure}
\centering
\includegraphics[width=\linewidth]{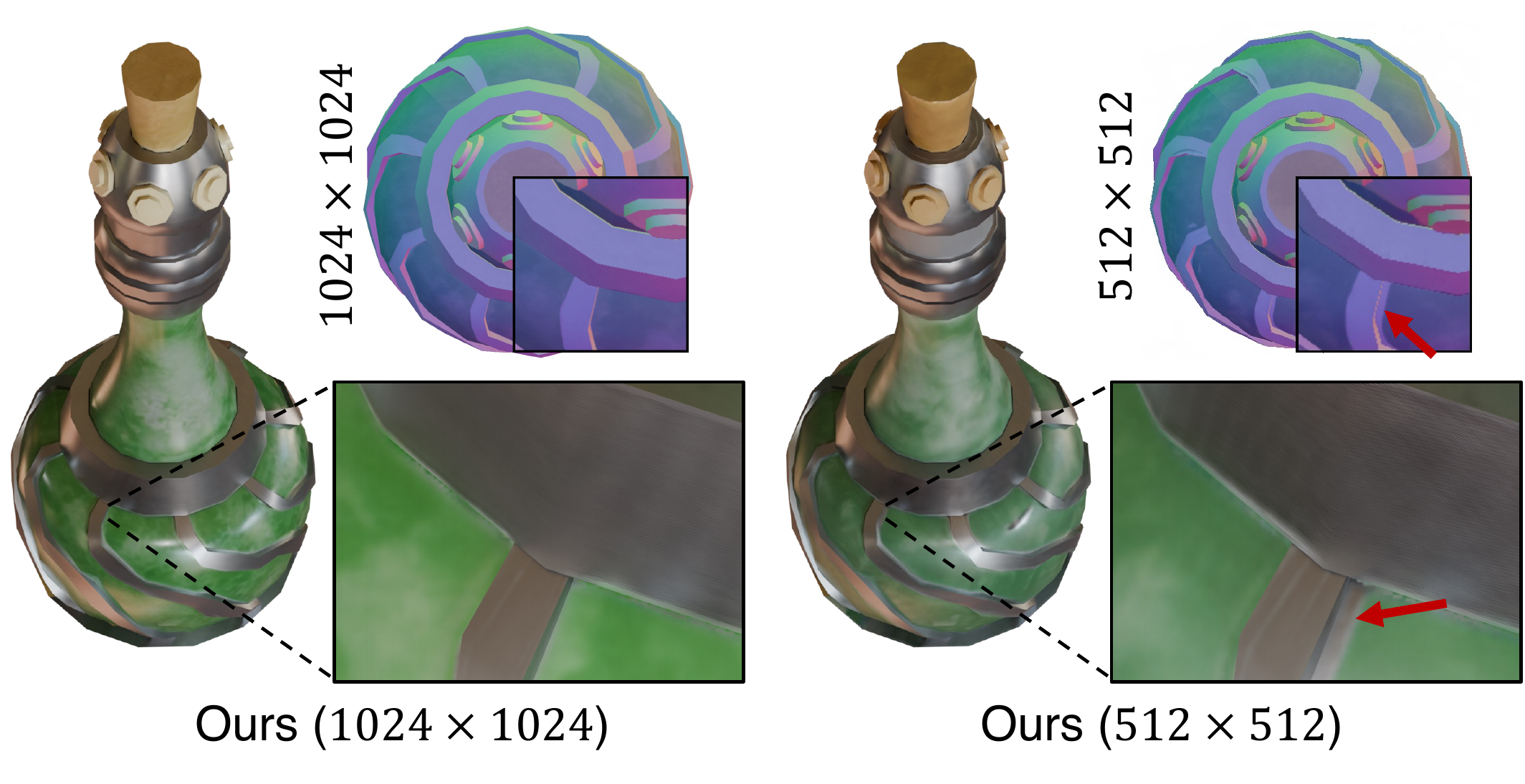}
\caption{\textbf{Generation with high resolution improves texture-geometry alignment.} The top right displays the blended results of normal and generated albedo to visualize the alignment.}
\label{fig:resolution} 
\end{figure}

\begin{table*}[t]
\centering
\caption{\textbf{Quantitative comparisons for single-view and multi-view settings.} Best results are marked as \textbf{1st} and \underline{2nd}.}
\resizebox{\textwidth}{!}{
\begin{tabular}{l
                !{\vrule width 0.7pt}
                cc c c cc
                !{\vrule width 0.7pt}
                cc c c cc}
\toprule
    & \multicolumn{6}{c!{\vrule width 0.7pt}}{\textbf{Single-view}} & \multicolumn{6}{c}{\textbf{Multi-view}}  \\

    \cmidrule(lr){2-7} \cmidrule(lr){8-13}
    Method
    & \multicolumn{2}{c}{Albedo} & Metallic & Roughness & \multicolumn{2}
    {c!{\vrule width 0.7pt}}{Rendering}
    & \multicolumn{2}{c}{Albedo} & Metallic & Roughness & \multicolumn{2}{c}{Rendering} \\

    \cmidrule(lr){2-3} \cmidrule(lr){4-4} \cmidrule(lr){5-5} \cmidrule(lr){6-7}
    \cmidrule(lr){8-9} \cmidrule(lr){10-10} \cmidrule(lr){11-11} \cmidrule(lr){12-13}
    & SSIM$\uparrow$ & PSNR$\uparrow$ & MSE$\downarrow$ & MSE$\downarrow$ & FID$\downarrow$ & LPIPS$\downarrow$
    & SSIM$\uparrow$ & PSNR$\uparrow$ & MSE$\downarrow$ & MSE$\downarrow$ & FID$\downarrow$ & LPIPS$\downarrow$ \\
\midrule
NvDiffRec~\cite{Munkberg_2022_CVPR}
    & 0.863 & 25.93 & 0.059 & 0.025 & 73.67 & 0.145 
    & 0.878 & 26.89 & 0.066 & 0.028 & 60.97 & 0.117 \\
Paint3D~\cite{zeng2024paint3d}
    & 0.877 & 24.66 & - & - & 60.54 & 0.166 
    & - & - & - & - & - & - \\
Material Anything~\cite{huang2025material}
    & 0.879 & 25.97 & 0.036 & 0.014 & 54.16 & 0.111 
  & 0.897 & 27.19 & 0.038 & 0.013 & 47.27 & 0.099 \\
MaterialMVP~\cite{he2025materialmvp}
    & 0.901 & 27.57 & 0.026 & 0.013 & 38.27 & 0.089
  & 0.902 & 27.61 & 0.026 & 0.013 & 38.00 & 0.088 \\
\midrule
Ours (stage1) + TexGEN~\cite{yu2024texgen}
    & 0.908 & 27.48 & - & - & 54.10 & 0.123 
    & \underline{0.927} & \underline{30.23} & - & - & 47.97 & 0.102 \\

Ours ($512 \times 512$)
    & \underline{0.920} & \underline{28.84} & \underline{0.025} & \underline{0.010} & \underline{31.64} & \underline{0.076}
  & \underline{0.927} & 29.92 & \underline{0.022} & \underline{0.010} & \underline{31.34} & \underline{0.068} \\

Ours ($1024 \times 1024$)
    & \bf 0.931 & \bf 29.89 & \bf 0.017 & \bf 0.007 & \bf 31.49 & \bf 0.066 
  & \bf 0.945 & \bf 32.10 & \bf 0.015 & \bf 0.008 & \bf 26.20 & \bf 0.052 \\
\bottomrule
\end{tabular}
}
\label{tab:quant_comp_grouped}
\end{table*}

\paragraph{Inference.}
The inference of the whole framework is performed on a single V100 GPU card.
In both prediction and generation stages, we select the predicted results of the first input image as the reference view. As high resolution improves alignment between geometry and texture in the generated results (Fig.~\ref{fig:resolution}), we perform inference at a resolution of $1024$ in our experiments. During generation, to fully leverage the resolution of each selected view, we adaptively rescale each camera intrinsic by evaluating the projection coverage of geometry surface within the view space, ensuring a high degree of coverage.
For meshes without UV mapping, we use Blender’s Smart UV Project to perform the UV unwrapping. For inference at $1024$ resolution, reconstructing materials for each object takes around $9$ to $23$ minutes (depending on the number of selected views $N$).

\begin{figure*}
\centering
\includegraphics[width=\linewidth]{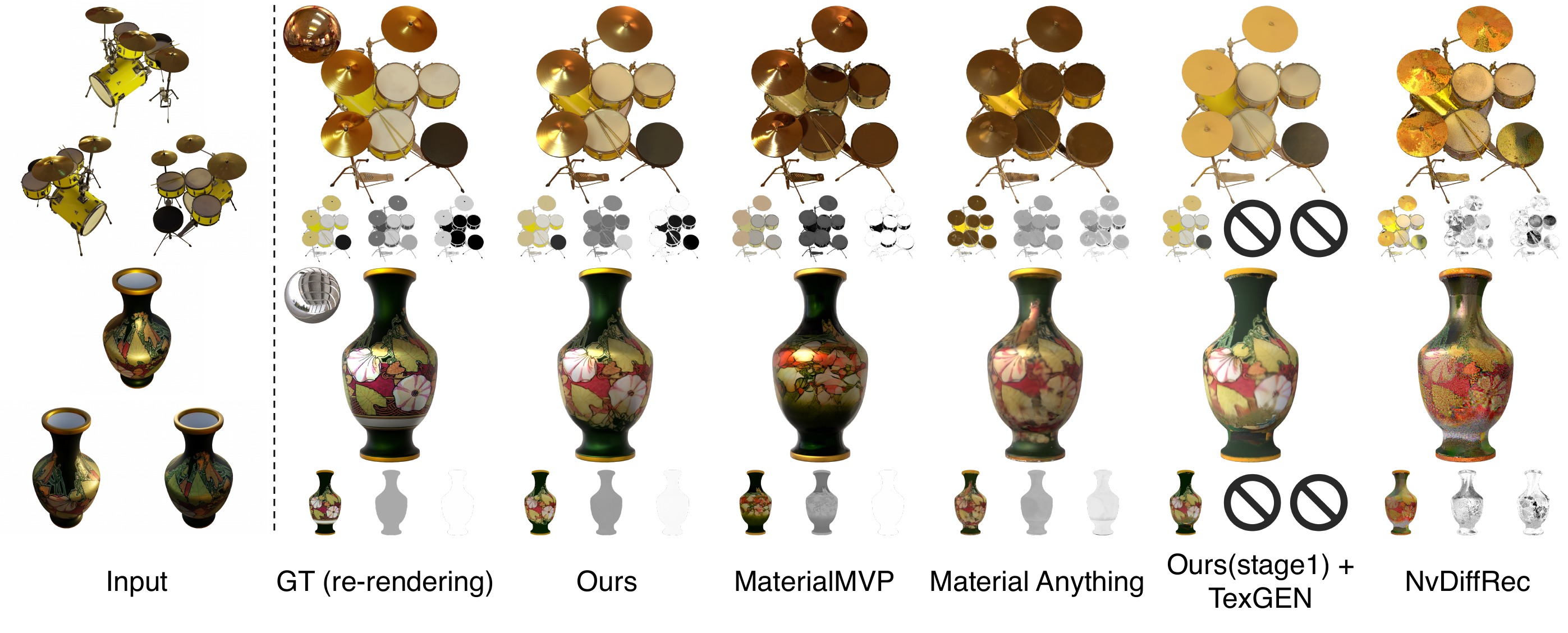}
\caption{\textbf{Qualitative comparison of multi-view input.} Our method recovers consistent materials from multi-view inputs. It preserves the detailed features of the input images, resulting in high-quality renderings. From left to right are albedo, roughness, and metallic.}
\label{fig:com-multi} 
\end{figure*}

\section{Experiments}
\label{sec:experiments}

\paragraph{Dataset.}
For testing, we select a subset of Objaverse~\cite{objaverse} comprising 100 objects, and render 9 views of each object with random environment maps. One or three views are used as inputs for single-view and multi-view comparisons, with the remaining views used for evaluation.

\paragraph{Baselines.}
To evaluate the effectiveness of our method, we select 5 representative approaches as baselines, including Material Anything~\cite{huang2025material}, MaterialMVP~\cite{he2025materialmvp}, Paint3D~\cite{zeng2024paint3d}, TexGEN~\cite{yu2024texgen}, and NvDiffRec~\cite{Munkberg_2022_CVPR}.
For Material Anything and MaterialMVP, we use our input images as the initial views and conditioning inputs, respectively. Paint3D only works in single-view setting, generating an albedo map conditioned on a single image. TexGEN generates textures conditioned on partial UV space albedo. Since this method is unable to process RGB inputs, we use the predicted albedo of our model (stage1) as its conditioning input. NvDiffRec is an inverse rendering method, we optimize each object using the input images for comparison.

\paragraph{Metrics.}
To evaluate the accuracy of the reconstructed materials, we compute the scale-invariant Peak Signal-to-Noise Ratio (PSNR) and the Structural Similarity Index Measure (SSIM)~\cite{zhou2004ssim} for albedo, and the Mean Square Error (MSE) for roughness and metallic.
To further assess the fidelity of generation, we calculate the Fr\'{e}chet Inception Distance (FID)~\cite{Seitzer2020FID} and the Learned Perceptual Image Patch Similarity (LPIPS)~\cite{zhang2018perceptual} on the rendered images.

\subsection{Comparisons}
\label{subsec:comparison}

We present quantitative comparisons in Tab.~\ref{tab:quant_comp_grouped}. Our method achieves more faithful albedo reconstruction, along with more accurate roughness and metallic estimations, resulting in superior FID and LPIPS scores on renderings. 
When multiple input views are available, the material reconstruction benefits substantially from leveraging the information contained in these views. Our framework is explicitly designed to exploit multi-view inputs, which leads to notable performance gains. Furthermore, higher-resolution inputs enable more precise alignment of fine details, and the substantial improvement in metrics at higher resolutions further validates the effectiveness of our design.

Qualitative comparisons are provided in Fig.~\ref{fig:com-single} and~\ref{fig:com-multi}, where the ground-truth bump map is applied to all results for better visualization.
As an inverse rendering framework, NvDiffRec~\cite{Munkberg_2022_CVPR} struggles to disentangle material and illumination with such sparse views, producing noisy renderings under novel illumination.
Paint3D~\cite{zeng2024paint3d} produces textures with baked shadows and reflections, as it is not explicitly trained for material prediction. 
TEXGen~\cite{yu2024texgen} leverages our material priors to generate high-quality textures. However, as the UV texture map is produced by the network, its limited resolution results in blurred outputs, such as the textures on the vase. Moreover, generation in the UV space tends to produce noticeable artifacts. For example, the red string on the blue bag is incorrectly generated as another color. 
MaterialMVP~\cite{he2025materialmvp} generates visually appealing objects and renderings, but fails to preserve the appearance of the input (such as the icon on the blue bag). In contrast, our method effectively preserves the texture features of the original image. 
Material Anything~\cite{huang2025material} also retains original textures by projecting estimated materials into UV space, yet its complex system leads to suboptimal material predictions, as discussed in Sec.~\ref{para:discuss}. By end-to-end training the two-stage framework, our method achieves accurate material prediction and consistent material generation, thereby resulting in realistic, semantically coherent outputs.

\begin{figure}
\centering
\includegraphics[width=\linewidth]{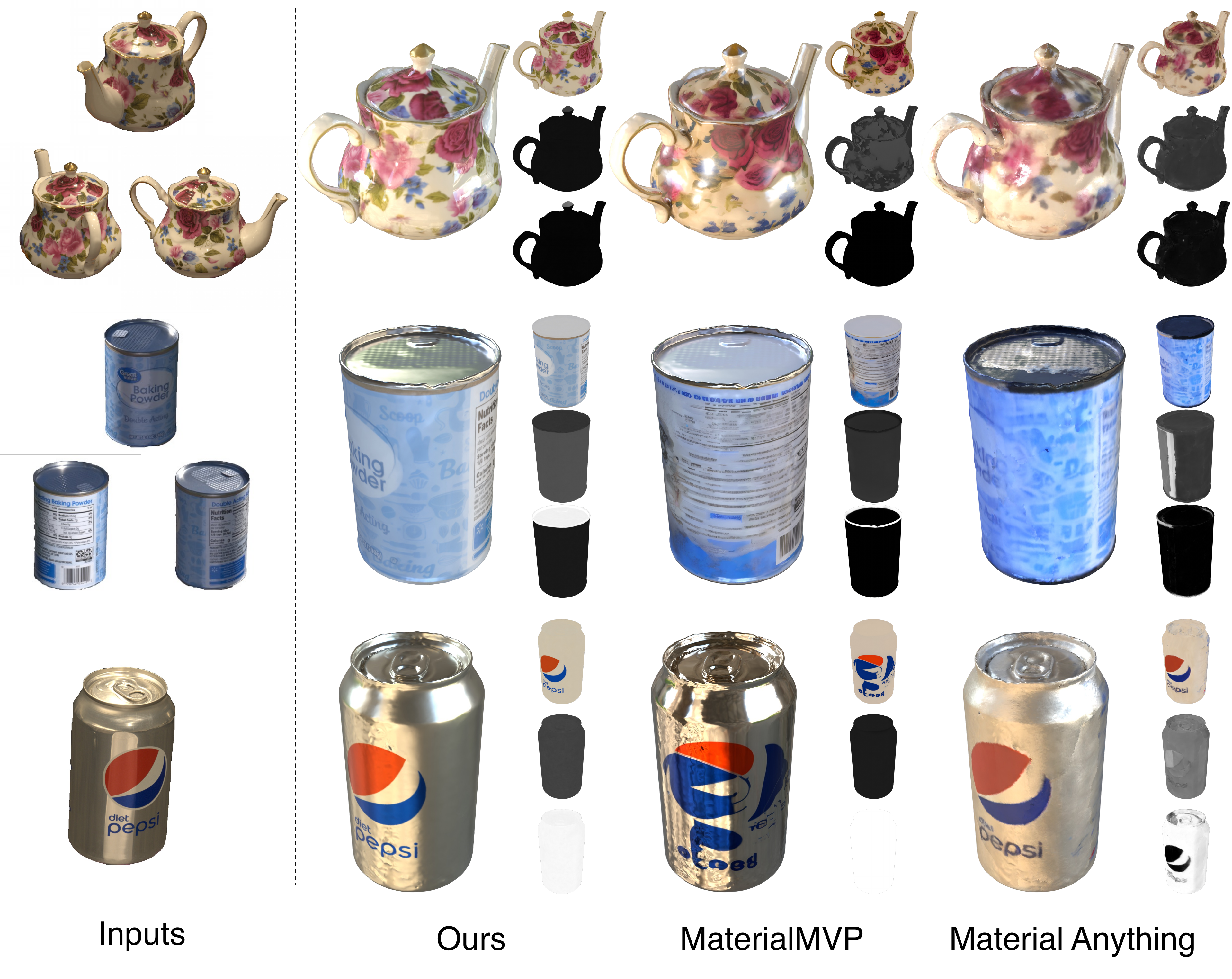}

\caption{\textbf{Qualitative comparison on real-world data Stanford-ORB~\cite{kuang2023stanfordorb}.} From top to bottom: albedo, roughness, and metallic.}
\label{fig:com_real_multi} 
\end{figure}

To further demonstrate the generalizability, we conduct additional experiments on real-world dataset Stanford-ORB~\cite{kuang2023stanfordorb} (Fig.~\ref{fig:com_real_multi}). Since most objects in this dataset exhibit strong view-dependent effects, we primarily focus on comparisons with MaterialMVP~\cite{he2025materialmvp} and Material Anything~\cite{huang2025material}. 
Similar to the results of synthetic data, MaterialMVP has difficulty maintaining texture structure, experiencing a loss of semantically meaningful information. 
Material Anything fails to produce consistent and seamless predictions (such as the seam on the Coke bottle) across views, especially when handling highly reflective objects, resulting in reduced rendering quality. 
Compared to these methods, our approach achieves the highest visual fidelity on real-world data, resulting in high-quality renderings.

\subsection{Ablation Study}

\begin{figure}
\centering
\includegraphics[width=\linewidth]{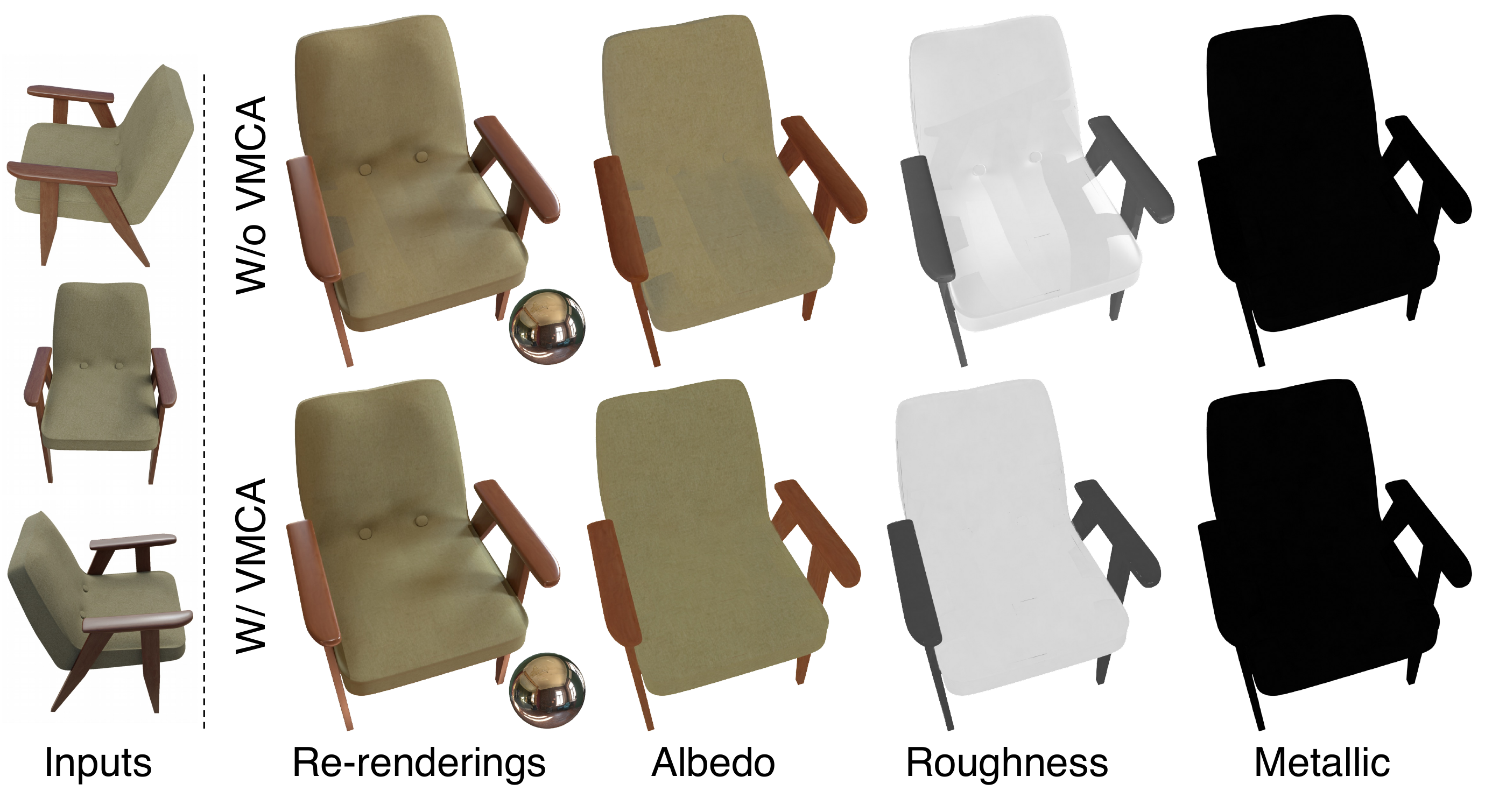}
\caption{\textbf{The effect of VMCA on reconstructed results.} VMCA improves the consistency of reconstructed results.}
\label{fig:vmca} 
\end{figure}

\paragraph{View-material Cross Attention.} 
We validate the effectiveness of VMCA through both quantitative (Tab.~\ref{tab:ablation_singview}) and qualitative comparisons (Fig.~\ref{fig:vmca_pred} and \ref{fig:vmca}). The prediction results shown in the Fig.~\ref{fig:vmca_pred} demonstrate that, without VMCA, the albedo and roughness of the barrel exhibit noticeable inconsistencies. For the final reconstruction, VMCA contributes to greater uniformity in the reconstructed materials, thereby resulting in higher-quality renderings (Fig.~\ref{fig:vmca}).

\begin{figure}
\centering
\includegraphics[width=\linewidth]{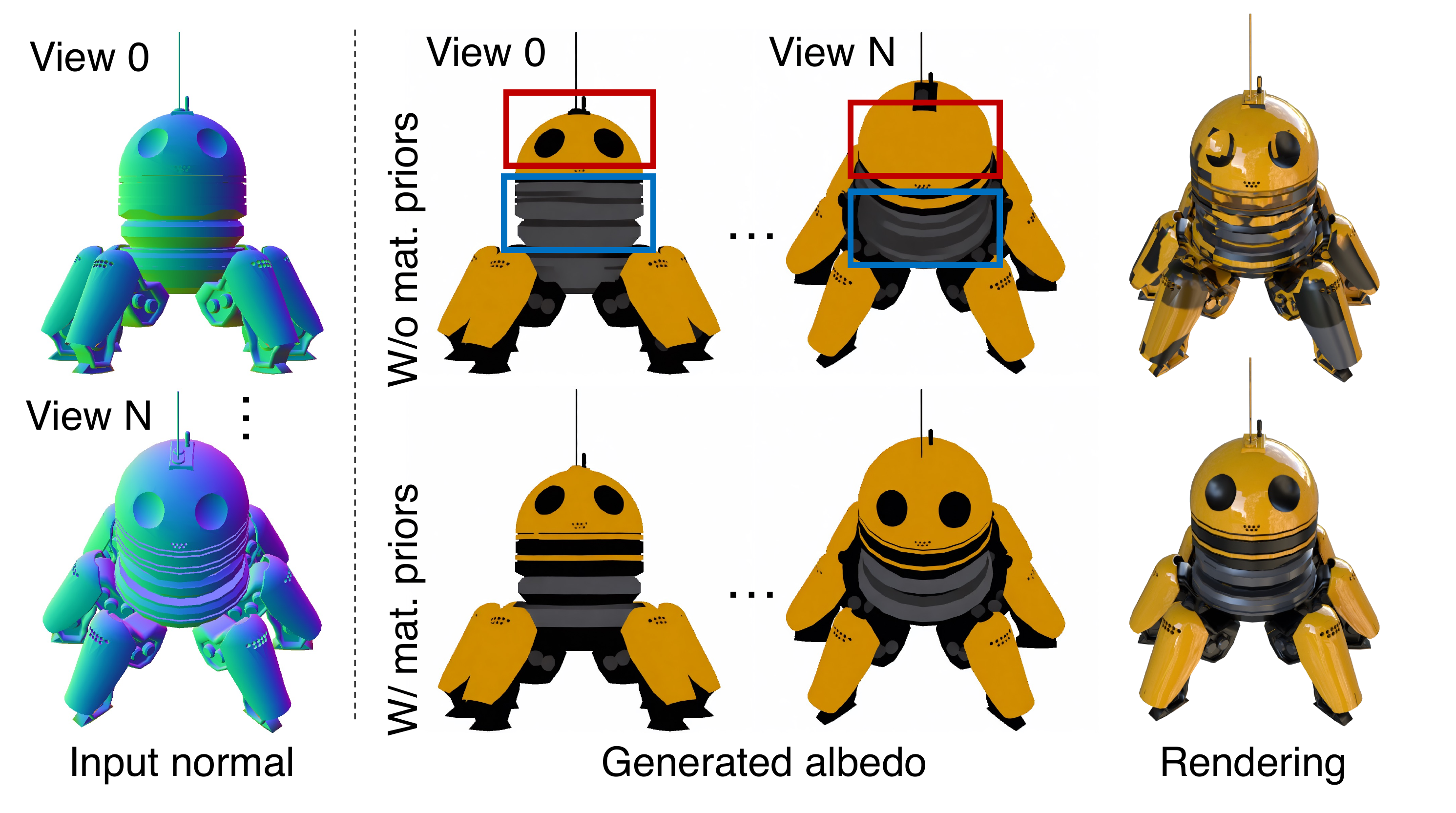}
\caption{\textbf{Material priors ablations.} Material priors ensure the consistency for progressive generation.}
\label{fig:mat} 
\end{figure}

\paragraph{Material Priors.} To evaluate the effectiveness of material priors during generation, we fill the material priors with black, thereby avoiding the network accessing any information from them. As shown in Fig.~\ref{fig:mat}, without material priors, the albedo generated based on geometric priors varies across views. These inconsistencies are blended in the same regions of UV space during baking, ultimately resulting in chaotic materials and reduced rendering quality (Tab.~\ref{tab:ablation_singview}).

\begin{figure}
\centering
\includegraphics[width=\linewidth]{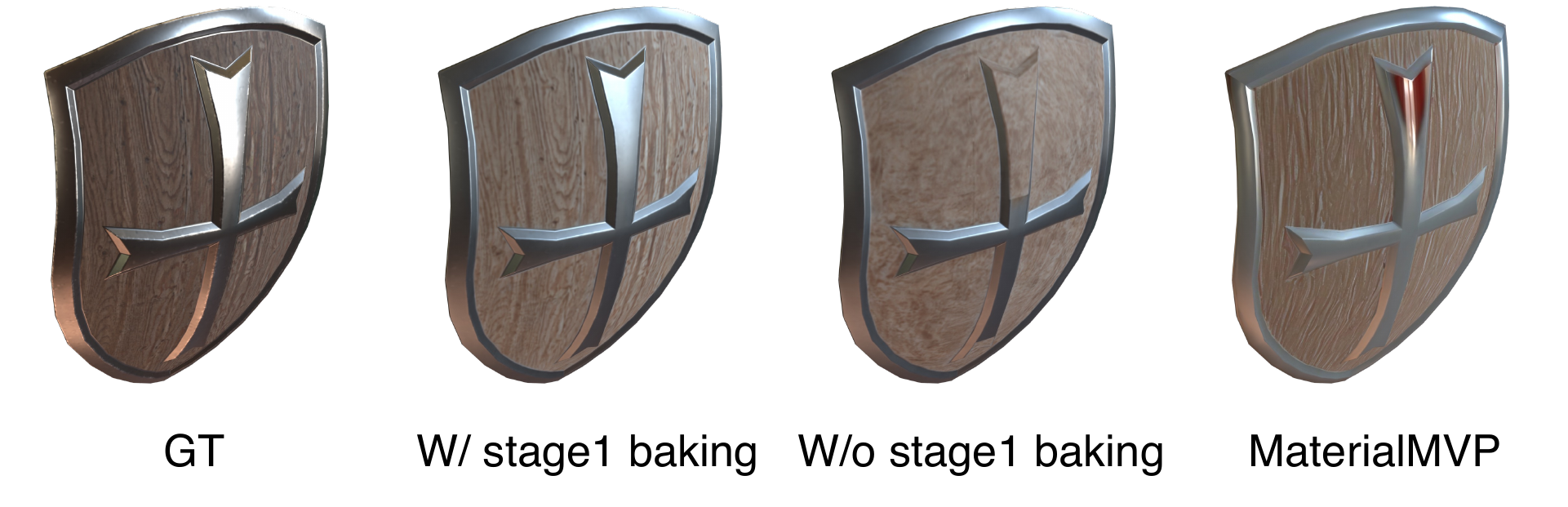}
\caption{\textbf{Stage1 baking improves the quality.}}
\label{fig:nobaking} 
\end{figure}

\paragraph{Texture Baking in Stage1.}
We evaluate the effect of stage1 texture baking on material reconstruction by skipping this step. In this case, similar to methods like MaterialMVP~\cite{he2025materialmvp} and Paint3D~\cite{zeng2024paint3d}, all materials are obtained through generation. Both quantitative (Tab.~\ref{tab:ablation_singview}) and qualitative (Fig.~\ref{fig:nobaking}) results indicate that achieving high-fidelity materials solely through generation remains a significant challenge. In contrast, generating materials based on the baked textures from stage1 fully leverages prior information, leading to higher-quality reconstructed results.

\begin{table}[t]
\centering
\caption{\textbf{Ablation study.} Experiments are conducted under a multi-view input setting. Best results are marked as \textbf{1st}.}

\resizebox{\linewidth}{!}{
\begin{tabular}{
 l 
 !{\vrule width 0.7pt}
 c  c  c  c  c  c
}
\toprule

\multicolumn{1}{c!{\vrule width 0.7pt}}{} &
\multicolumn{2}{c}{Albedo} &
\multicolumn{1}{c}{Metallic} &
\multicolumn{1}{c}{Roughness} &
\multicolumn{2}{c}{Rendering} \\
\cmidrule(lr){2-3} \cmidrule(lr){4-4} \cmidrule(lr){5-5} \cmidrule(lr){6-7}
 & SSIM$\uparrow$ & PSNR$\uparrow$ & MSE$\downarrow$ & MSE$\downarrow$ & FID$\downarrow$ & LPIPS$\downarrow$ \\
\midrule
W/o VMCA             & 0.941 & 31.78 & 0.016 & \bf 0.007 & 28.58 & 0.055 \\
W/o mat. priors      & 0.929 & 29.85 & 0.018 & 0.009 & 34.93 & 0.071 \\
W/o stage1 baking    & 0.917 & 28.14 & 0.021 & 0.009 & 42.40 & 0.091 \\
Ours (full)          & \bf 0.945 & \bf 32.10 & \bf 0.015 & 0.008 & \bf 26.20 & \bf 0.052 \\
\bottomrule
\end{tabular}
}
\label{tab:ablation_singview}
\end{table}

\section{Conclusion}
\label{sec:conclusion}
\ttt\ achieves high-quality material recovery through a two-stage reconstruction approach, making it a practical framework. Compared with other methods, this framework demonstrates high scalability and can be deployed by end-to-end optimization of a single diffusion model. 
However, as with other material prediction methods, the inherent ambiguity of material decomposition may cause scaled color for the predicted albedo, and objects with strong self-occlusion may require a greater number of views for generation. We leave these limitations for future work.
\clearpage

{
    \small
    \bibliographystyle{ieeenat_fullname}
    \bibliography{main}
}


\end{document}